\documentstyle[12pt,a4,graphicx]{article}

\newcommand{\be}{\begin{equation}}
\newcommand{\ee}{\end{equation}}
\newcommand{\ba}{\begin{eqnarray}}
\newcommand{\ea}{\end{eqnarray}}
\newcommand{\dis}{\displaystyle}
\newcommand{\re}{\mbox{Re}\,}
\newcommand{\im}{\mbox{Im}\,}
\newcommand{\tr}{\mathrm{tr}}

\begin{document}
\begin{titlepage}
\begin{flushright}
{UG-FT-121/00}\\
\end{flushright}
\vspace{2cm}
\begin{center}

{\large\bf   Chiral Limit Prediction for $\varepsilon_K'/\varepsilon_K$
at NLO in $1/N_c$ \footnote{Work supported
in part by NFR, Sweden, by CICYT, Spain, 
by Junta de Andaluc\'{\i}a, (Grant No. FQM-101),
and by the European Union TMR Network $EURODAPHNE$ (Contract
No. ERBFMX-CT98-0169).
Invited talk at ``High Energy Euroconference on Quantum Chromodynamics
 (QCD '00)'', 6-13 July 2000, Montpellier, France.}}\\
\vfill
{\bf Johan Bijnens$^{a)}$  and Joaquim Prades$^{b)}$}\\[0.5cm]
$^{a)}$ Department of Theortical Physics 2, Lund University\\
S\"olvegatan 14A, S 22362 Lund, Sweden.\\[0.5cm]

$^{b)}$ Departamento de
 F\'{\i}sica Te\'orica y del Cosmos, Universidad de Granada\\
Campus de Fuente Nueva, E-18002 Granada, Spain.\\[0.5cm]
\end{center}
\vfill
\begin{abstract}
\noindent
We report on a calculation of $\varepsilon_K'/\varepsilon_K$
at next-to-leading in the $1/N_c$ expansion and to lowest order
in Chiral Perturbation Theory. We  discuss the short-distance
matching, the scale and scheme dependence as well as
 the long-distance
short-distance matching. We include the two known chiral corrections
to our result and discuss further order $p^4$ corrections to it.
\end{abstract}
\vfill
September 2000
\end{titlepage}

\section{Introduction}
Recently, 
direct CP violation in the Kaon system has been
unambiguously established by the KTeV \cite{KTEV}
experiment at Fermilab and by the NA48 \cite{NA48} experiment at CERN.
Their results together with the previous 
 NA31 \cite{NA31} and  E731 \cite{E731} measurements
produce the present world average
\ba
{\re} \left(\varepsilon_K'/\varepsilon_K\right) \, 
&=& (19.3\pm2.4) \cdot 10^{-4}  \, .
\ea
Further reduction of the statistical error
to the order of $1 \cdot 10^{-4}$ is expected. 

A lot of effort has been put in the theoretical
side  to get a Standard Model prediction
for this quantity in the last twenty five  years.
Recent reviews and predictions are 
\cite{Munich,Roma,PR,Trieste,Dortmund,Narison,PPI00}.
 Here, we would like to report on a calculation
of this quantity in the chiral limit and next-to-leading (NLO) order
in  $1/N_c$  \cite{epsprime}. We also discuss how
it changes when the  known chiral corrections, i.e. final state interactions
(FSI) and $\pi_0-\eta$ mixing are included.
Comments on the different approaches to obtain
the contributions from $Q_6$ and $Q_8$ to $\varepsilon_K'$
will also be given in the Summary.

Direct CP-violation in  $K\to\pi\pi$ decays amplitudes
is parameterized by
\ba
\frac{\varepsilon_K'}{\varepsilon_K}&=&
\frac{1}{\sqrt 2}\left[ \frac{A\left[ K_L \to (\pi\pi)_{I=2}\right]}
{A\left[K_L \to (\pi\pi)_{I=0}\right]} 
\right. \nonumber \\
&-& \left. 
 \frac{A\left[ K_S \to (\pi\pi)_{I=2}\right]}
{A\left[K_S \to (\pi\pi)_{I=0}\right]} \right] \, .
\ea
We want  here to predict $K\to\pi\pi$
 at NLO  in $1/N_c$ and to lowest order in the chiral expansion.

In the isospin symmetry limit, $K\to\pi\pi$ invariant amplitudes
can be decomposed into definite isospin quantum numbers as
$[A\equiv-i T]$
\ba
\label{isospin}
i \, A[K^0\to \pi^0\pi^0] &\equiv& {\frac{a_0}{\sqrt 3}} \, 
 e^{i\delta_0}
-\frac{ 2 \, a_2}{\sqrt 6} \, 
a_2 \, e^{i\delta_2} \, , \nonumber \\
i \, A[K^0\to \pi^+\pi^-] &\equiv& {\frac{a_0}{\sqrt 3}} \, 
 e^{i\delta_0}
+\frac{a_2}{\sqrt 6} \, e^{i\delta_2}\,  
\ea
with $\delta_0$ and $\delta_2$ the
FSI phases. 

To lowest order in Chiral Perturbation Theory (CHPT), i.e. 
order $e^0 p^2$ and $e^2 p^0$, strong and electromagnetic
interactions between  $\pi$, $K$, $\eta$ and external sources are 
described by
\be
\label{lagstrong}
{\cal L}^{(2)} =
\frac{F_0^2}{4}\tr\left(u_\mu u^\mu+\chi_+\right)
+ e^2 \tilde C_2 \tr\left(Q U Q U^\dagger\right)
\ee
with $U = u^2 = \exp(\lambda^a\pi^a/F_0)$ and 
$u_\mu = i u^\dagger  (D_\mu U) u^\dagger$.  
$\lambda^a$ are the Gell-Mann matrices and the $\pi^a$
are the pseudoscalar-mesons $\pi$, $K$, and $\eta$.
$Q=\mbox{diag}(2/3,-1/3,-1/3)$ is the light-quark-charge matrix and
$\chi_+ = u^\dagger\chi u^\dagger + u \chi^\dagger u$ and
$\chi = 2 B_0 \, \mbox{diag}(m_u,m_d,m_s)$ collects the light-quark 
masses. To this order, $F_\pi=F_0=$ 87 MeV is the pion decay coupling 
constant.  Introductions to CHPT can be found in Refs.
\cite{CHPTlectures,TASI}.

To the same order in CHPT, the chiral Lagrangian
describing $|\Delta S|=1$ is
\ba
\label{lagdS1}
{\cal L}^{(2)}_{|\Delta S|=1}&=&
C F_0^6 e^2 G_E \,  \tr\left(\Delta_{32}\tilde{Q}\right)
\nonumber \\ &&\hskip-1.5cm+
 C F_0^4 \Bigg[G_8\tr\left(\Delta_{32}u_\mu u^\mu\right)
+G_8^\prime\tr\left(\Delta_{32}\chi_+\right) 
\nonumber \\
&&\hskip-1.5cm+G_{27}t^{ij,kl}\tr\left(\Delta_{ij}U_\mu\right)
  \tr\left(\Delta_{kl}u^\mu\right)\Bigg] +\mbox{h.c.}
\ea
with $\Delta_{ij} = u\lambda_{ij}u^\dagger$,
$(\lambda_{ij})_{ab} = \delta_{ia}\delta_{jb}$,
$\tilde{Q}=u^\dagger Q u$; 
\ba
C&=& -\frac{3}{5} \frac{G_F}{\sqrt 2} \, V_{ud} V_{us}^* \approx
-1.06 \cdot 10^{-6} \, {\rm GeV}^{-2} \, . 
\ea
The SU(3) $\times$ SU(3) tensor
$t^{ij,kl}$ can be found in \cite{kptokpp}.
Using this Lagrangian, 
\ba
a_2 & = & \frac{\sqrt 3}{9} \, C F_0\left[10 G_{27} \, 
(m_K^2-m_\pi^2) - 6 e^2 G_E F_0^2 \right]
\nonumber\\
a_0 & = &\frac{\sqrt 6}{9} \, C F_0\left[\left(9G_8+G_{27}\right) 
(m_K^2-m_\pi^2) \right. 
\nonumber \\ &-& \left.  6 e^2 G_E F_0^2 \right]
\ea
and $\delta_0=\delta_2=0$.
In the presence of CP-violation, the couplings $G_8$, $G_{27}$, and
$G_E$ get an imaginary part. In the Standard Model,
$\im  G_{27}$ vanishes and
$\im  G_8$ and $\im  G_E$ are proportional
to $\im  \tau$ with
 $\tau \equiv -\lambda_t /\lambda_u$ and $\lambda_i\equiv
V_{id} V_{is}^*$ and where $V_{ij}$ are Cabibbo-Kobayashi-Maskawa 
(CKM) matrix elements. 

\section{Short-Distance Scheme and Scale Dependence}
\label{short}

Strangeness changing transitions 
happen in the Standard Model by
 the  exchange of one
$W$-boson.  This fact implies that {\em all} physics between
0 and $\infty$  has to be taken into account
when calculating Kaon to pions amplitudes.
In particular, this implies the intervention
of  strong interactions at all energies,
i.e. from the perturbative region to the 
non-perturbative one.

The two very different scales involved in $K\to\pi\pi$,
i.e.  the $W$-mass and the Kaon mass make 
effective field theory methods very useful. This is standard
and we outline the steps needed.
 Below the $W$-mass, the effective action 
$\Gamma_{\Delta S=1}$  is obtained by integrating out 
the heavy particles, top, $Z$, and $W$-bosons and using 
the operator product expansion (OPE).
The leading contributions consist of four-quark operators
and higher dimensional operators are suppressed
by $M_W^2$. Using the renormalization group equations,
$\Gamma_{\Delta S=1}$ is brought down to some perturbative scale
below the charm quark mass. Perturbative matching and OPE
is used at thresholds of the successive heavy particles 
which are integrated out.
This full process implies several choices of short-distance schemes,
regulators, and operator basis. Of course, physical matrix elements 
cannot depend on these choices. 
 
Explicit calculations have been performed including gluonic and 
electromagnetic  Penguin-diagrams to one-loop in \cite{one-loop} 
and to two-loops  in \cite{two-loop} in two schemes.
These are the naive dimensional regularization (NDR) 
scheme and the   't Hooft-Veltman (HV) scheme.
The choice of the operator basis and other choices
like infrared regulators can be found in those references. 

The Standard Model $\Gamma_{\Delta S=1}$ effective action
at scales $\nu$ below the charm quark mass, is \cite{Buras}
\ba
\label{MSbar}
\Gamma_{\Delta S=1}&\sim& 
{\dis \sum_{i=1}^{10}}\, C_i(\nu)\,  \int {\rm d}^4 x \, Q_i(x)
+ {\rm h.c.}
\ea
where $C_i = z_i + \tau  y_i$ are Wilson coefficients
and $Q_i(x)$ are four-quark operators.
 At this level, we have resummed the large logarithms 
 $[\alpha_s(\nu)  \log(\nu/M_W)]^n$
and $\alpha_s [\alpha_s(\nu) \log(\nu/M_W)]^n$ 
to all orders in $n$.

At low energies, it is more convenient to describe 
the $\Delta S=1$ transitions with an effective action
$\Gamma_{\Delta S=1}^{LD}$ which uses hadrons, constituent quarks,
or other objects to describe the relevant degrees of 
freedom. A regularization scheme like an
Euclidean  cut-off
which separates long-distance physics 
from the short-distance physics which is integrated out and 
working in four dimensions is also more practical, as well as
 another operator basis, for instance 
the color singlet Fierzed one. 
The effective action  $\Gamma_{\Delta S=1}^{LD}$ 
depends on all these choices and in particular on the scale $\mu_c$
introduced to regulate the divergences generated
analogously as (\ref{MSbar}) depends on the scale $\nu$.
It also depends on effective couplings
$g_i$ analogously to the Wilson coefficients in (\ref{MSbar}).
As usual, matching conditions have to be set between the 
two effective 
field theories. This is done by requiring that $S$-matrix elements
of asymptotic states are the same at some perturbative scale
\ba
\label{matching}
\langle 2 | \Gamma_{\Delta S=1}^{LD} | 1 \rangle &=&
\langle 2 | \Gamma_{\Delta S=1} | 1 \rangle \, .
\ea
Both sides are separately 
scale and scheme independent
 and therefore the short-distance 
scale and scheme dependences are consistently treated.
This matching  is done at the perturbative level using 
the OPE in QCD.
 Notice that even if the regulator chosen
is the same in both sides,  there can be finite terms appearing 
in the matching. The matching conditions 
 fix {\em analytically} the short-distance behavior of the
couplings $g_i$
\ba
g_i(\mu_c,\cdots)&=& {\cal F}(C_i(\nu), \alpha_s(\nu), \cdots) \, .
\ea
This was done explicitly in \cite{scheme} for
$\Delta S=2$ transitions and used in  \cite{epsprime}
for $\Delta S=1$ transitions.
To the best of our knowledge, these are the first
places were  the short-distance
scheme dependence is analytically cancelled
in the final matrix element at next-to-leading
order in the $1/N_c$ expansion.
Notice that in the whole process we never required
the matching of scales in two different regularization schemes
as is sometimes stated in the literature 
for this type of $1/N_c$ calculations.

\subsection{The heavy $X$-bosons Method}

We find it convenient to use
an effective field theory of heavy color-singlet
$X$-bosons coupled to QCD currents and densitities
 as argued in \cite{scheme,Bardeen,BK,DeltaI=1/2}.
One advantage of this is that two-quark currents are unambiguously
identified and that QCD densities are much easier to match
than four-quark operators. 
E. g., the operator
\ba
Q_1(x)&=&  \left[\overline s \gamma_\mu (1-\gamma_5)  d \right](x) \,
 \left[\overline u \gamma_\mu (1-\gamma_5)  u \right](x)
\nonumber
\ea
is reproduced by the effective action
\ba
\label{Q1}
\Gamma_{X}&\equiv& g_1(\mu_c,\cdots) \int {\rm d}^4 y 
\, X_1^\mu \left\{ \left[\overline s \gamma_\mu (1-\gamma_5)  
d \right](x) \right. \nonumber \\ &+& \left. 
 \left[\overline u \gamma_\mu (1-\gamma_5)  u \right](x) \right\}
\ea
We use an Euclidean cut-off at $\mu_c$ 
in four dimensions as regulator of the UV divergences, 
with this the $X$-boson effective action is completely specificied.
This regulator is very convenient to separate long- from
short-distance physics. 
 In (\ref{Q1}), the higher than $\mu_c$ degrees of freedom
of quarks and gluons have been integrated out.

The complete $X$-boson effective action for the 
Standard Model $\Delta S=1$ 
transitions can be found in \cite{epsprime}.
We are now ready to calculate consistently $\Delta S=1$ Green functions
with the $X$-boson effective theory.

\section{Long-Distance--Short-Distance Matching}
\label{long}

We study the two-point $\Delta S=1$ Green function
$$
\label{greens}
\Pi(q^2)\equiv i \int {\rm d}^4 x \, e^{iq \cdot x}
\langle 0 | T(P_i^\dagger(0) P_j(x) e^{i \Gamma_{X}}|0\rangle \, 
$$
where $P_i$ are pseudoscalar sources
with the quantum numbers to describe $K\to  \pi$
amplitudes. 

After reducing the external legs one gets
$K\to\pi$ off-shell amplitudes.
Taylor expanding them in external momentum and $\pi$, and $K$ masses
one can obtain the couplings of the CHPT Lagrangian
\cite{kptokpp,BDSPW}. Then one predicts $K\to\pi\pi$
at a given order. This procedure is unambiguous.
One can use also $K\to\pi$ and $K\to$ vacuum transitions
form lattice simulations to predict  the lowest
order CHPT couplings \cite{kptokpp,BDSPW,GP}.

Up to know we have not used any $1/N_c$ argument.
We count the $g_i$ couplings as effectively order 1 since 
they contain  the large logarithms
 of the running between $M_W$ and $\mu_c$
even though some of them start at NLO order in $1/N_c$.
At leading order in the $1/N_c$ expansion the contribution  to the 
Green's function (\ref{greens}) is factorizable.
It only involves two-point functions.
At this order the result is model independent
and the scale and scheme dependence is exactly taken into account.
However, we know that the factorizable contribution
fails to reproduce $\re a_0$  by a factor around 6 
and $\re a_2$ by a factor around 1/2.

At the same order, $Q_6$ contributes to
$G_8$ proportional to
\ba
|g_6|^2 \sim 
C_6(\nu) \langle 0 | \overline q q | 0 \rangle^2(\nu) \,
 L_5(\hat \mu) 
\ea
and $Q_8$ to $e^2 G_E$
proportional to
\ba
|g_8|^2 \sim 
C_8(\nu) \langle 0 | \overline q q | 0 \rangle^2(\nu)  \, .
\ea
To leading order 
in $1/N_c$, the  scale dependence of the Wilson coefficients
cancels exactly the one from the quark condensate \cite{Q6BG,Q6EdR}.
 The scale $\hat \mu$ dependence of $L_5$ 
requires NLO in $1/N_c$ contributions to cancel. 
In fact, at this order there also 
appears an IR divergence in the factorizable 
contribution of $Q_6$ to $G_8$ 
that necessarily requires non-factorizable
contributions to cancel \cite{DeltaI=1/2}.

The non-factorizable contribution is NLO in the $1/N_c$
expansion and involves the integration of four-point functions
$\Pi_{P_iP_jJ_aJ_b}$ over the $X$-boson momentum
$r_E$ that flows through the currents/densities $J_a$ and $J_b$.
 It can be schematically written as
\ba
\Pi(q^2) \sim \int \frac{{\rm d}^4 r_E}{(2\pi)^4}\, 
 \Pi_{P_iP_jJ_aJ_b} (q_E, r_E) \, .
\ea
We separate long- from short-distance physics with
an Euclidean cut-off $\mu$ in $r_E$. The short distance
part of the integral from $\mu$ up to $\infty$ 
can be treated at NLO within OPE QCD using
a cut-off regulator if $\mu$ is high enough.

It has been emphasized that 
dimension eight operators may be numerically important
when the cut-off scale $\mu$ is of the order of 1 GeV 
\cite{BK,CDG00}. 
This issue can be treated in a straightforward way in our approach.
We have to use the OPE in the short-distance part
up to the required dimensions.
In fact, those contributions are under control at NLO in $1/N_c$ 
using factorized matrix elements.
We stopped at dimension six operators for the present results,
we plan to check the effect of dimension eight operators elsewhere.

Up to here, there is no model dependence in our evaluation
of $K\to \pi$ amplitudes at NLO in $1/N_c$ within QCD.

What remains is the long distance part from 0 up to $\mu$.
For very small values of $\mu$ one can use CHPT 
and still the result is model independent. However, 
CHPT at order $p^4$ starts not to be enough 
already at relatively small values of $\mu$, i.e. $\sim$ 400 MeV.
To match with the short-distance part, we need clearly to go beyond
 this energy. The first step is to use a good hadronic model
for intermediate energies. The model we used is the ENJL model
described in \cite{ENJL}. It has several good features 
-it includes CHPT to order $p^4$, for instance-
and also some drawbacks as explained in \cite{scheme}.
Work is in progress to implement the large $N_c$ constraints
along the lines of \cite{PPR98}.
There is very interesting work on a different approach
to the calculation of weak matrix elements within the
$1/N_c$ expansion in \cite{BK,LMD}. It has started to produce the first 
results for three- and four-point Green's functions.
In particular the determination of the $B_K$ parameter
in the chiral limit is clean \cite{BKLMD}.
Their result agrees well with the chiral limit calculation
in \cite{scheme,BK} done within our present approach. 

\section{$\varepsilon'_K$ in the Chiral Limit}
\label{results}

To a very good approximation \cite{TASI,Buras},
\ba
\label{epsilonprime}
\left| \varepsilon_K'\right| &\simeq&
\frac{1}{ \, \sqrt 2} \frac{\re a_2}{\re a_0}\, 
\left\{ -\frac{\im a_0}{\re a_0}+\frac{\im a_2}
{\re a_2}\right\}\, .
\ea
The isospin amplitudes $a_I$ are defined in (\ref{isospin}).
The lowest order CHPT values for $\re a_0$ and $\re a_2$ can be
obtained from a fit to $K\to \pi\pi$ and $K\to \pi\pi\pi$
amplitudes up to order $p^4$ in \cite{KMW91}.

\begin{figure}[thb]
\begin{center}
\includegraphics[height=0.5\textwidth,angle=270]{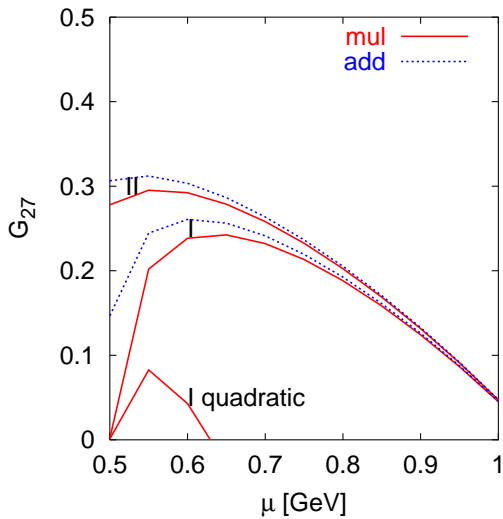}
\caption{\label{G27} We plot the result for $G_{27}$.
Curves I and II are different ways of doing the two-loop
$\alpha_s$ running and {\it mul} and {\it add}
are the resummation on the Wilson coefficients
is done either multiplicatively or additively.
The curve {\it quadratic} is the result of references
\cite{Dortmund,BBG}.}
\end{center}
\end{figure}

\begin{figure}[thb]
\begin{center}
\includegraphics[height=0.5\textwidth,angle=270]{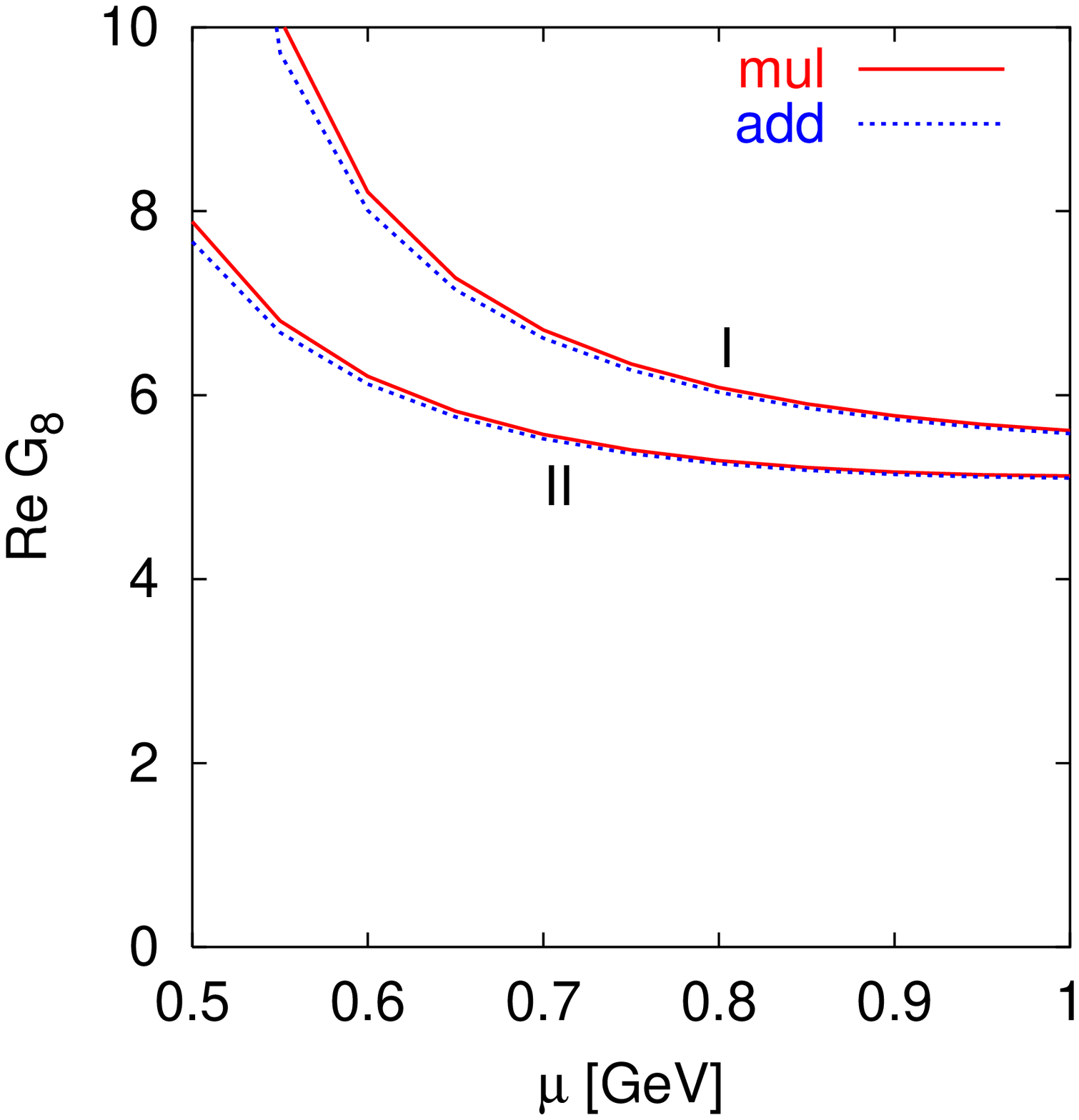}
\caption{\label{G8real}  We plot the result for $\re G_8$.
The labels are like in Figure \ref{G27}.}
\end{center}
\end{figure}

Our results in \cite{DeltaI=1/2} reproduce the
value for $\re  a_0$ finding a nice matching of
short-and long-distances. For the coupling $G_{27}$,
which dominates $\re a_2$, we don't find  such good stability though 
the behaviour is much better than the quadratic
divergence found in \cite{Dortmund,BBG}.
 In Figure \ref{G27} we give $G_{27}$ and in Figure \ref{G8real}
we present $\re G_8$.

The $\Delta I=1/2$ enhancement is reproduced within  40 \%.
However, due to the present lack of a
complete understanding of $\re a_2$,
we prefer to use the {\it experimental} values for 
the lowest order CHPT amplitudes $\re a_I$ 
\cite{KMW91} to predict $\varepsilon_K'$.

At lowest order in CHPT, 
the imaginary part of $G_8$ is almost all from
the $Q_6$ operator with very small contributions
from $Q_3$, $Q_4$, and $Q_5$. Then to a very good
approximation
\ba
\im  G_8 &\simeq& - \frac{80}{3} \, 
\im \tau \, y_6(\nu) \nonumber \\
&\times&
\frac{\langle 0 | \overline q q | 0 \rangle^2(\nu)}{F_0^6}  
\, L_5(\hat \mu) \, B_6(\hat \mu, \nu).
\ea
At large $N_c$, $B_6(\hat \mu, \nu)=1$.
At the same order, the imaginary part of $e^2 G_E$ is
almost all from the $Q_8$ operator with a very small
contribution from $Q_7$. Then to a very good approximation
\ba
\im (e^2 G_E) &\simeq&  - 5 \, \im \tau \,  y_8(\nu) 
\nonumber \\
&\times &
\frac{\langle 0 | \overline q q | 0 \rangle^2(\nu)}{F_0^6} 
\, B_8(\nu)
\ea
and $B_8(\nu)=1$ at large $N_c$.
With the present values of the CKM matrix elements
\be
\im \tau = - 6.72 \cdot 10^{-4} \,.
\ee

We use  the chiral limit $\overline{MS}$ quark-condensate \cite{BPR95}, 
\ba
B_0( 1 \rm GeV) &\equiv&-
\frac{\langle 0 | \overline q q | 0  \rangle (1 \rm GeV)}{F_0^2}
\nonumber \\ &=& (1.75\pm0.40) \, {\rm GeV} \, .
\ea
Its  leading scale dependence is analytically
canceled by the one in the Wilson coefficients 
$C_6$ and $C_8$ \cite{Q6BG,Q6EdR}. The uncertainty 
induced in $\varepsilon_K'$ by $B_0$
is around  40 \% and often not included
in error estimates. 
Incidentally, higher order corrections to
the contribution of $Q_6$ to $G_8$ do not change
the chiral limit condensate into the strange quark condensate as
is commonly used \cite{DeltaI=1/2}. 

There is no NLO in $1/N_c$ correction to $e^2 G_E$ from $Q_8$
in nonet symmetry. This is a model independent result.
In the real world, the U(1)$_A$ anomaly
gives a mass to the pseudoscalar
singlet field  and octet symmetry is a good symmetry. 
Therefore, there are $1/N_c^2$ corrections to this result 
which we intend to study within 
our $1/N_c$ approach elsewhere  \cite{BGP} using 
data to obtain $B_8$ \cite{Narison,PPR98,DG00}
as well as a lowest meson dominance model.
They will give the  size of the U(1)$_A$ anomaly corrections
to our NLO in $1/N_c$ results. The results in  \cite{Narison,DG00}
studied the $B_8$ parameter with the additional
assumption that dimension six operators dominate the OPE.

Our values of the relevant bag parameters 
in the NDR scheme at 2 GeV are in Table \ref{tabBi}.
\begin{table*}[htb]
\caption{\label{tabBi} Results for $B_{6}(\nu)^{(1/2)NDR}$,
$B_{7}(\nu)^{(3/2)NDR}$ and $B_{8}(\nu)^{(3/2)NDR}$ at $\nu=2$~GeV.
 The ${\overline {MS}}$
 value ${\overline m_s}(2 {\rm GeV})= (119\pm 12)$ MeV was used
 in \cite{Narison} and 
for rescaling the results in \cite{DG00,LMDQ7}.}
\newcommand{\m}{\hphantom{$-$}}
\newcommand{\cc}[1]{\multicolumn{1}{c}{#1}}
\renewcommand{\arraystretch}{1.2} 
\begin{tabular}{@{}llll}
\hline
Method & $B_{6}^{(1/2)NDR}(2 {\rm GeV})$ & 
$B_{7}^{(3/2)NDR}(2{\rm GeV })$ & $B_{8}^{(3/2)NDR} (2 {\rm GeV})$\\
\hline
NLO $1/N_c$ [$\chi$ limit] \cite{epsprime}& 
$2.5\pm0.4$ & $0.8\pm 0.1$ & $1.35\pm0.20$\\
LMD \cite{LMDQ7} &-- & 0.94 & -- \\
Dispersive  \cite{DG00} &-- & $0.78\pm0.17$ & $1.6\pm0.4$\\
QCD Sum Rules \cite{Narison} & $1.0\pm 0.5$ & $0.7\pm 0.2$ 
& $1.70\pm0.39$\\
CHPT NLO $1/N_c$ \cite{Dortmund} & $1.5\sim1.7$ & $-0.1\sim0.09$
 & $0.4\sim0.7$\\
Lattice \cite{latticeB7} &-- & $0.5\sim0.8$ & $0.7\sim1.1$\\
Chiral Quark Model \cite{Trieste} & $1.2\sim1.7$ & $\sim0.9$ & $\sim0.9$ \\
FSI Omn\`es \cite{PPI00} & $1.55$ & -- & $0.93$ \\
\hline
\end{tabular}\\[2pt]
\end{table*}

Details and the input parameters
of the determination of $\varepsilon_K'/\varepsilon_K$ 
 are in \cite{epsprime}. Here we only give the results
and main conclusions.
To lowest order in $1/N_c$ and in the chiral limit, 
we get
\ba
\label{LOchi}
\left|\frac{\varepsilon_K'}{\varepsilon_K}\right|_\chi
&=& \left( 33 - 16 \right) \cdot 10^{-4} 
=  (17 \pm 7 \pm 6)\cdot 10^{-4} \nonumber \\
&=& (17 \pm 9)\cdot 10^{-4}
\ea
The first error is from the uncertainty in the value
of the quark condensate and the second is a
typical  1/3 on the factorizable contributions.
We cannot give here an error
estimate due to the non-inclusion of 
non-factorizable $1/N_c$ corrections 
since they are a new type of contributions.

\begin{figure}[thb]
\begin{center}
\includegraphics[height=0.6\textwidth,angle=270]{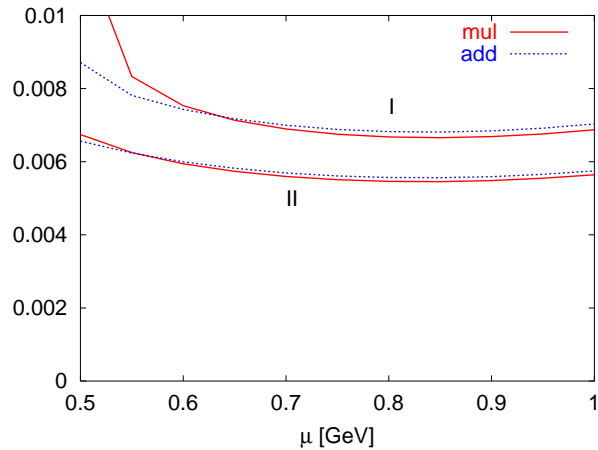}
\caption{\label{epspeps} We plot our result
for $\im G_8$ at NLO in $1/N_c$.
The labels of the curves are as for Figure \ref{G8real}.}
\end{center}
\end{figure}

Including our calculated NLO in $1/N_c$ non-factorizable 
contributions we get
\ba
\label{NLOchi}
\left|\frac{\varepsilon_K'}{\varepsilon_K}\right|_\chi
&=& \left( 83 - 23 \right) \cdot 10^{-4} 
=  (60 \pm 24\pm 20 )\cdot 10^{-4} \nonumber \\
&=& (60 \pm 30 )\cdot 10^{-4}  
\ea
again in the chiral limit.
The first error is once more due to the uncertainty in the 
quark condensate. 
The second one is an estimate of the model uncertainty, i.e. around 
1/3  of the factorizable contribution plus 40 \%
of the non-factorizable one added quadratically.
Notice that there is a factor larger than three between
(\ref{LOchi}) and (\ref{NLOchi}).
This result is plotted in Figure \ref{epspeps}
where one can appreciate the quality of the matching.

\begin{figure}[thb]
\begin{center}
\includegraphics[height=0.5\textwidth,angle=270]{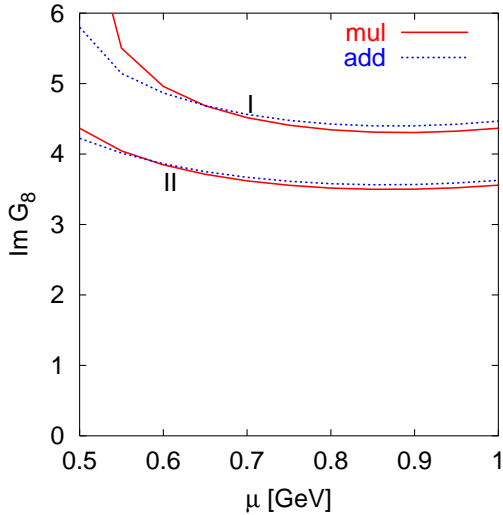}
\caption{\label{G8imag} We plot our result
for $\im G_8$ at NLO in $1/N_c$.
The labels of the curves are as for Figure \ref{G8real}.}
\end{center}
\end{figure}

\begin{figure}[thb]
\begin{center}
\includegraphics[height=0.5\textwidth,angle=270]{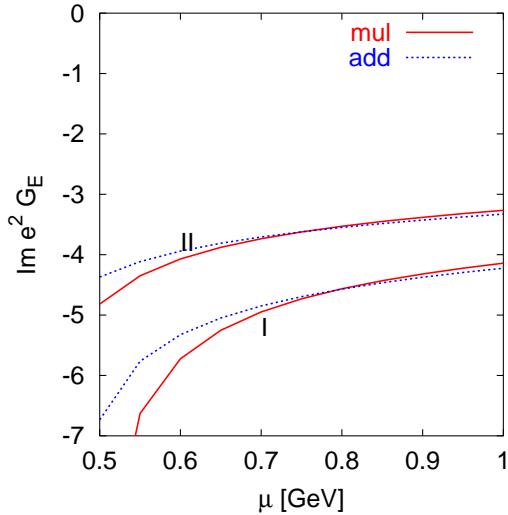}
\caption{\label{GEimag} We plot our result
for $\im G_E$ at NLO in $1/N_c$.
The labels of the curves are as for Figure \ref{G8real}.}
\end{center}
\end{figure}

To get the result in (\ref{NLOchi}) we have used our chiral
limit NLO  in $1/N_c$ determinations
of $\im G_8$ which we  show in Figure \ref{G8imag}
and of $\im G_E$ which is in Figure \ref{GEimag}.

\section{Higher Order CHPT Corrections}
\label{higher}

The r\^ole of final state interactions 
in the standard \cite{Munich,Roma} predictions
of $\varepsilon_K'/\varepsilon_K$ has been recently
clarified in \cite{FSI}. See also \cite{moreFSI}.

We have taken a different strategy to predict
$\varepsilon_K'$. We start with Equation (\ref{epsilonprime}).  
The ratio  $\im a_I / \re a_I$ has no final state interactions
to all orders. One of the main problems at present of
  $K\to\pi\pi$  lattice calculations is that they cannot
 include  FSI. There are very interesting recent developments
towards a full calculation within lattice QCD including FSI,
see \cite{LL00} and references therein for previous work.

 It would be very interesting to see how much
of the ratios $\im a_I /\re a_I$ could still be obtained from
lattice simulations {\it without} FSI.
One could use 
CHPT to eliminate  analytically the FSI contributions
for both the real and imaginary parts.

The only place where FSI are present is in the normalization
factor $\re a_2 /\re a_0$, but one can take as a first
step its experimentally known value.
In fact, lattice calculations can provide us
with an estimate of $\varepsilon_K'/\varepsilon_K$
to lowest order CHPT with $K \to \pi$ and $K\to$ vacuum 
results \cite{epsprime}. This would be very interesting.
Of course, one also wants
to predict the $\Delta I=1/2$ enhancement 
ratio $\re a_2 /\re a_0$ but 
$K\to\pi$ calculations 
look more straightforward in the lattice at present \cite{GP}.

About isospin breaking due to $m_u\neq m_d$,
only  effects due to $\pi^0-\eta$ mixing
\cite{Q6BG,pi0eta} are under control. 
They have recently calculated to
CHPT order $p^4$ \cite{pi0eta}. Though $\pi^0-\eta$
mixing is all from $m_u\neq m_d$ isospin breaking 
at lowest CHPT order, this is not true at higher orders.
In fact, due to the smallness of the order $p^4$ term
in  the $\pi^0-\eta$ mixing
contribution, it could be that there are  other $m_u-m_d$
effects equally important 
at the same order, so that a full $p^4$ calculation
is mandatory \cite{isospin,pi0eta}.
This effect adds  a contribution  to  $\im \, a_2$  which is
parameterized usually as
\ba
\frac{\left[\im \, a_{2}\right]_{IB}}{\re \, a_2} &=& \Omega_{IB} \, 
\frac{\im \, a_0}{\re \, a_0} \, . 
\ea
We will use  the value $\Omega_{IB} = 0.16\pm 0.03$ \cite{pi0eta}

Other purely real $p^4$ corrections even in the isospin
limit are mostly unknown. They have been taken
into account within the approach in \cite{Trieste}
and partially  in \cite{Dortmund}.

Electromagnetic corrections have been considered in \cite{Q6EdR,em}.
Very little is known at present of the size of the order
$p^4$ contributions and beyond.
Obviously there is more work to be done in this direction.

To lowest order in $1/N_c$ but including FSI
and $\pi^0-\eta$ chiral corrections -notice that these
two corrections are actually NLO in $1/N_c$- we get
\ba
\label{LOnonchi}
\left|\frac{\varepsilon_K'}{\varepsilon_K}\right|
&=& \left( 24 - 15.5\right) \cdot 10^{-4} 
= (8.5 \pm 3.5 \pm 3)
\cdot 10^{-4}  \nonumber \\ 
&=& (8.5 \pm 4.6)\cdot 10^{-4} \, .
\ea
The errors correspond to the same discussion of the
chiral limit results (\ref{LOchi}) and in particular
do not include the error of the non-included non-factorizable
contributions. This result is model independent.

Notice that small changes in
the isospin zero or in the isospin two contributions
are extremely amplified, so that a 20 \% correction 
to both but in opposite  directions can translate into a factor two
in the value of $\varepsilon_K'$.

Including our calculated NLO in $1/N_c$ non-factorizable 
contributions, we get
\ba
\label{NLOnonchi}
\left|\frac{\varepsilon_K'}{\varepsilon_K}\right|
&=& \left( 59.6 - 25.6 \right) \cdot 10^{-4}   
=  (34 \pm 14 \pm 11 )\cdot 10^{-4} \nonumber \\
&=& (34 \pm 18 )\cdot 10^{-4} \, .
\ea
Notice that the two known chiral corrections 
reduce the value of $\varepsilon_K'$.

\section{Summary}

We have reported on a calculation of $\varepsilon_K'/\varepsilon_K$
at next-to-leading in a $1/N_c$ expansion and
in the chiral limit \cite{epsprime}. 
Emphasis has been given  to discuss the short-distance scheme and 
scale dependence matching in Section \ref{short}
and  to the long-distance short-distance matching in Section
\ref{long}. 

We have also given the two model
independent results at leading order in the $1/N_c$ expansion
and in the chiral limit in (\ref{LOchi}) and including the two known chiral
corrections, namely FSI and $\pi^0-\eta$ mixing in (\ref{LOnonchi}).

We also have given the main result of our work which is
the value of $\varepsilon_K'/\varepsilon_K$ in the chiral limit
and at next-to-leading order in a $1/N_c$ expansion
(\ref{NLOchi}). Including again the two known higher order
CHPT corrections  to this result, we obtain our final 
number for $\varepsilon_K'/\varepsilon_K$ in (\ref{NLOnonchi}).
Notice however, as discussed in Section \ref{higher},
that most of the higher order CHPT corrections are actually
unknown and as remarked in several recent work
they could affect numerically the 
Standard Model value of $\varepsilon_K'/\varepsilon_K$ 
significantly \cite{epsprime,isospin,em}. 

We would like to discuss the  determinations 
of the two main bag parameters for $\varepsilon_K'$
in Table \ref{tabBi}.
About $B_6$, which is the most uncertain, 
at present, there are only two estimates of
the non-factorizable corrections not proportional
to quark masses to  this important parameter.
They are the results  in \cite{Dortmund}
and in \cite{epsprime}.

The Dortmund group \cite{Dortmund} 
has included non-factorizable 
corrections to $B_6$ in the chiral limit.
For that, they use  CHPT to order $p^4$ and
find a quadratic dependence in the cut-off scale. 
They have also included partially higher order CHPT
corrections to that result.

To the best of our knowledge, the result in 
\cite{epsprime},
is the first one that includes NLO in $1/N_c$
non-factorizable corrections to $B_6$ in the chiral limit
and find a logarithmic matching with the Wilson coefficients
see Figure \ref{epspeps} for $\varepsilon_K'/\varepsilon_K$
and Figure \ref{G8imag} for the imaginary part of $G_8$.

The Trieste group \cite{Trieste} uses $B_6=1$ 
in the chiral limit and all the difference between 
the factorizable result and their final number
quoted in the Table \ref{tabBi} are higher order CHPT corrections. 
They have not calculated 
non-factorizable corrections in the chiral limit.

Notice that the calculation of $B_6$ in \cite{Narison}
only includes the factorizable contributions.

In \cite{PPI00}, the leading
order in $1/N_c$ values of $B_6=1$ 
and $B_8=1$ are corrected 
with  $\pi \pi$ final state
interactions using the Omn\`es solution.
In our opinion, this is at present the best way of taking into account
within CHPT those important corrections.

For $Q_8$ we go at NLO in $1/N_c$ but in the chiral limit, and
we find that, as explained in Section \ref{results} and
\cite{epsprime}, 
\be
B_8^{NDR}(2 {\rm GeV})_\chi = 1.35 \pm 0.20
\ee
is a model independent result.
There are however $1/N_c^2$ corrections induced by the U(1)$_A$
anomaly. These effects were estimated in \cite{Narison,DG00}
with the additional assumption that  dimension six
operators dominate the OPE in this case. We plan to investigate
this issue within our approach \cite{BGP}.

Again, only NLO corrections proportional to quark masses are
included in the result of the Trieste group to $B_8$
\cite{Trieste}. 
The results from the lattice \cite{latticeB7} do not include FSI but we
don't known how much do they 
include of other higher order chiral corrections.

Let us conclude that a lot of advances have been done
towards the prediction of $\varepsilon_K'/\varepsilon_K$ 
within the Standard Model.
For us, there are two main questions:
which is the value of $B_6$  in the chiral limit
and which are the non-FSI chiral corrections to it.
This work tries to answer the first of these questions.
There is going on a lot of activity to clarify these
issues by several groups and we foresee a very
exciting nearby future.

\section*{Acknowledgements}

It is a pleasure to thank Stephan Narison for the invitation
to this very enjoyable conference. J.P. also thanks the hospitality
 of  the Centre de Physique Th\'eorique, CNRS-Luminy at Marseille where
part of his work was done and of the 
Department of Theoretical Physics at Lund University where
 this work was written.

\end{document}